\documentclass[reprint,prr,longbibliography,nofootinbib,amsmath,amssymb,aps]{revtex4-2}
\usepackage{graphicx}% Include figure files
\usepackage{dcolumn}% Align table columns on decimal point
\usepackage{bm}% bold math
\usepackage{mathtools}
\usepackage{csquotes}
\usepackage{hyperref}% add hypertext capabilities
\usepackage{float}
\usepackage{bigints}
\usepackage{color}

\begin{document}

% \preprint{APS/123-QED}

\newcommand{\physrep}{{Phys. Rep.}}

\newcommand{\AAA}{\boldsymbol{A}}
\newcommand{\BB}{\boldsymbol{B}}
\newcommand{\JJ}{\boldsymbol{J}}
\newcommand{\EE}{\boldsymbol{E}}
\newcommand{\UU}{\boldsymbol{U}}
\newcommand{\kk}{\boldsymbol{k}}
\newcommand{\xx}{\boldsymbol{x}}
\newcommand{\rr}{\boldsymbol{r}}
\newcommand{\Bnabla}{\boldsymbol{\nabla}}
\newcommand{\ii}{\mathrm{i}}
\newcommand{\bra}[1]{\langle #1\rangle}
\newcommand{\bbra}[1]{\left\langle #1\right\rangle}
\newcommand{\eqss}[2]{(\ref{#1})--(\ref{#2})}
\newcommand{\EEq}[1]{Equation~(\ref{#1})}
\newcommand{\Eq}[1]{Eq.~(\ref{#1})}
\newcommand{\Eqs}[2]{Eqs.~(\ref{#1}) and~(\ref{#2})}
\newcommand{\Eqss}[2]{Eqs.~(\ref{#1})--(\ref{#2})}
\def\EM{E_{\rm M}}
\def\EA{E_{\rm A}}
\def\HM{H_{\rm M}}
\def\Lu{\mbox{\rm Lu}}
\def\MATINS{\texttt{MATINS}}
\def \clara#1{{\color{blue}CD: #1}}
\def \jose#1{{\color{blue}JP: #1}}

\title{Magnetar field dynamics shaped by chiral anomalies and helicity}

\author{Clara Dehman$^{1}$}
\email{clara.dehman@ua.es}
\author{Jos\'e A. Pons$^{1}$}

\affiliation{$^{1}$ Departament de Física Aplicada, Universitat d'Alacant, 03690 Alicante, Spain}

\newcommand{\jcap}{{JCAP}}
\newcommand{\aap}{{A\&A}}
\newcommand{\mnras}{{MNRAS}}
\newcommand{\apjl}{{ApJL,}}
\newcommand{\apjs}{{ApJS,}}
\newcommand{\ssr}{Space Sci. Rev.,}

\begin{abstract}
The chiral magnetic effect (CME)---a macroscopic manifestation of the quantum chiral anomaly---induces currents along magnetic field lines, facilitating mutual conversion between chiral asymmetry and magnetic helicity. Although the finite electron mass suppresses chiral asymmetry through spin-flip processes, we demonstrate that the CME effectively shapes magnetar field evolution. Magnetic helicity acts as a persistent internal source of chiral asymmetry, which mediates the redistribution of magnetic energy across spatial scales, without requiring an external energy source. 
Our three-dimensional magneto-thermal simulations of the neutron star crust reveal a novel mechanism that significantly reconfigures the magnetic field inherited at birth, amplifying both toroidal and poloidal large-scale dipolar components (crucial for spin-down) to strengths of $>10^{14}$~G within just a century, at the expense of small-scale structures.
This astrophysical application of the CME, distinct and complementary to conventional hydrodynamic dynamo models, offers an innovative framework for understanding magnetar field dynamics and provides a transformative solution to the origin of their exceptionally strong, large-scale fields.
\end{abstract}

\maketitle

The origin and evolution of magnetic fields in neutron stars (NSs) -- especially magnetars, the class with the strongest known magnetic fields -- remain subjects of active debate \cite{Mereghetti2015,esposito2021}. It is generally agreed that the large-scale dipolar fields observed in these objects, as inferred from spin periods and period derivatives, cannot be solely attributed to fossil fields inherited from their progenitor stars. Consequently, turbulent dynamo amplification is often invoked to explain their extreme strengths \cite{balbus1991,obergaulinger2014,raynaud2020,reboul2021,aloy2021}. Yet despite extensive study, critical questions persist, particularly regarding how magnetic energy transfers to larger scales, leaving the origin of magnetar fields unresolved.

Beyond macroscopic hydrodynamic processes such as dynamos and turbulence, quantum field theory establishes a fundamental microscopic connection between chirality and helicity through the chiral anomaly. This phenomenon facilitates a two-way transfer between fermionic chiral asymmetry and magnetic helicity. Recent research has increasingly focused on linking the generation of large-scale magnetic fields to chiral asymmetries produced during core-collapse supernovae and the proto-NS phase \cite{masada2018,matsumoto2022}. However, significant challenges persist, most notably the efficiency of spin-flip scattering processes induced by the finite electron mass, raising doubts about the CME’s relevance in these environments (see Kamada et al.~\cite{kamada2023} for a review on the CME in different scenarios). The rapid, temperature-dependent spin-flip reactions reduce the chiral imbalance, constraining the effectiveness of the chiral instability mechanism \cite{grabowska2015,sigl2016}.

The concept that a pre-existing helical magnetic field at short wavelengths could experience an inverse-like cascade driven by chiral asymmetry, ultimately generating large-scale magnetic fields, has been proposed in the context of the early Universe \cite{boyarsky2012,tashiro2012,pavlovic2017}, but has been considered far less often in stellar environments \cite{kaplan2017,matsumoto2022}. In this work, unlike previous studies, we incorporate chiral effects into the study of NSs, explicitly accounting for spin-flip scattering processes in our numerical simulations. We explore the long-term interaction between magnetic helicity and chiral asymmetry during the early stages of NS evolution. The simulations reveal that, while spin-flip processes significantly suppress chiral effects, residual asymmetries can persist for hundreds of years, enabling sustained energy transfer from small to large scales. In this context, the chiral anomaly acts as a catalyst, potentially reorganizing turbulent magnetic fields---originally generated by dynamo processes---to form a coherent, large-scale dipolar field.

The mechanism operates as follows: a small but sustained imbalance between left- and right-handed electrons generates an electric current parallel to the magnetic field---an effect known as the Adler-Bell-Jackiw anomaly \cite{adler1969,bell1969}. This imbalance is quantified by the chiral chemical potential $\mu_5 \equiv \mu_R - \mu_L$\footnote{Note that definitions of $\mu_5$ vary in the literature, sometimes differing by a sign \cite{sigl2016} or a factor of 2 \cite{kaplan2017}, depending on the source.}, where $\mu_R$ and $\mu_L$ are the chemical potentials of right- and left-handed electrons, respectively. A non-vanishing $\mu_5$ implies that magnetic field evolution in NS interiors must include an effective chiral degree of freedom, despite being within the Standard Model. When $\mu_5 \neq 0$, Maxwell's equations acquire an additional current term \cite{vilenkin1980}:
\begin{equation}
    \JJ_5 = \frac{\alpha \mu_5}{\pi \hbar} \BB,
    \label{eq: J5}
\end{equation}
where $\alpha = e^2/\hbar c$ is the fine structure constant, $e$ is the fundamental charge, $\hbar$ is the reduced Planck constant, and $c$ is the speed of light. We use Gaussian units throughout the paper.

In an MHD context, the chiral current ($\JJ_5$) acts analogously to a dynamo, amplifying magnetic fields by drawing on the energy stored in the chiral chemical potential. Notably, it can have either sign and reverse the process -- tapping magnetic energy to generate chirality.
In this letter, we examine this microphysical mechanism as an alternative -- or more precisely, a complement -- to classical hydrodynamical dynamos in explaining magnetar fields.

NSs exhibit a complex internal structure with multiple fluid components, each possessing distinct hydrodynamical velocities. In the outer crust, a rigid ionic lattice severely limits ion mobility, allowing only electrons to flow freely and sustain the currents essential for magnetic field evolution. The inner crust introduces added complexity with superfluid neutrons, which partially decouple from the nuclear lattice and act as a neutral fluid component. Complexity peaks in the NS core, where coexisting superfluid neutrons and superconducting protons necessitate a sophisticated multi-fluid framework. Given this complexity, magnetic field evolution is typically studied using region-specific approximations.
In this context, we focus on the crust, where only (slowly drifting) electrons are mobile within the solid ionic lattice, making the e-MHD approximation applicable \cite{pons2019}. The magnetic field evolution is then governed by the induction equation derived from Faraday’s law:
\begin{equation}
    \frac{\partial \BB}{\partial t} = - c \Bnabla \times \EE, 
    \label{eq: faraday law}
\end{equation}
where the electric field accounts for Ohmic dissipation, Hall drift, and the new chiral magnetic contribution:
\begin{eqnarray}
  c \EE =    \eta \left( \Bnabla \times \BB - k_5  \BB \right)  
    + f_h \left(\Bnabla \times \BB \right)\times \BB .  
    \label{Efield}
\end{eqnarray}
Here, $\eta = c^2/4\pi \sigma_e$ is the magnetic diffusivity, where $\sigma_e$ is the electrical conductivity;
$k_5 = 4 \alpha \mu_5/\hbar c$ is the chiral wavenumber; and $f_h = c/4\pi e n_e$ is the Hall prefactor, with $n_e$ the electron number density. 

Magnetic field evolution must be coupled to the evolution equation for the chiral number density $n_5 \equiv n_R - n_L$ \cite{adler1969,grabowska2015,rogachevskii2017,kamada2023}, which includes both source and sink terms:
\begin{equation}
    \frac{\partial n_5}{\partial t} =  \frac{2 \alpha }{\pi \hbar} \EE \cdot \BB + {n_e} \Gamma_w^\mathrm{eff} - n_5 \Gamma_f.
    \label{eq: n5}
\end{equation}
Here, the reaction rate $\Gamma_f$ accounts for spin-flip interactions and acts as a sink term, while $\Gamma_w^\mathrm{eff}$ represents the effective weak reaction rate \cite{epstein1981} and serves as a source.
The $\EE \cdot \BB$ term governs the coupling between the chiral density and the electromagnetic field: twisting or untwisting magnetic field lines alters the net chirality in the system, acting as either a source or a sink depending on its sign.

%%%%%%%%%%%%%%%%%%%%%%%%%%%%%%%%%%%%%%%%%
% Generalized helicity balance equation
%%%%%%%%%%%%%%%%%%%%%%%%%%%%%%%%%%%%%%%%%
\EEq{eq: n5} should be viewed alongside the time evolution of the magnetic helicity, which takes the form \cite{biskamp1997,boyarsky2012}:
 \begin{eqnarray}
\frac{\partial (\AAA \cdot \BB)}{\partial t} 
&=& - 2 c  \EE \cdot \BB -  c \Bnabla \cdot \left( \EE \times \AAA\right) .
 \label{eq: chirality and magnetic helcity}
\end{eqnarray}
The two equations, when combined and integrated over a volume, yield a generalized helicity balance law:
 \begin{equation}
    \frac{d}{dt} \left( Q_5 + \frac{\alpha }{\pi \hbar c} \chi_m  \right) + {\Gamma_5} = 0.
    \label{eq: modified helicity conservation}
\end{equation}
Here, $Q_5 = \int n_5 \, dV$ is the total axial charge, $\chi_m = \int \AAA \cdot \BB \, dV$ is the total magnetic helicity, and ${\Gamma_5} = \int n_5 \Gamma_f dV $ is the average spin-flip rate. 
Total helicity is not strictly conserved due to the sink term $\Gamma_5$. We neglect both the helicity flux across the boundary $\left(\propto \EE \times \AAA \right)$ and the weak interaction processes ($\Gamma_w^\mathrm{eff} \ll \Gamma_f$), as their contributions are marginal. Weak interactions could, however, transiently enhance the effect if the NS is out of chemical equilibrium. For simplicity, we set $\Gamma_w^\mathrm{eff} = 0$ hereafter, since its typical values ($10^{-2}- 10^2$\,s$^{-1}$) are many orders of magnitude smaller than the spin-flip rates (\Eq{eq: gamma flip}).

Concerning the spin-flip reaction rate, electrons interact via Rutherford, electron-electron, or Compton scattering. In the core, Rutherford scattering dominates \cite{dvornikov2015,sigl2016}, while in the crust, electron-nucleus scattering prevails. For degenerate electrons, the flip rate is given by \cite{haensel2007}:
\begin{equation}
\Gamma_f = \frac{4 \alpha }{3\pi \sigma_e}
\label{eq: gamma flip} \frac{m_e^2 c^4}{\hbar^2},
\end{equation}  
which is on the order of $10^{15}-10^{17}$ s$^{-1}$ under typical neutron star conditions during the first few centuries.
Since $\sigma_e$ depends on temperature, so does $\Gamma_f$, underscoring the need to consider the coupled evolution of temperature and magnetic field in simulations.

Given that all reaction rates are much faster than typical astrophysical timescales, we treat all quantities in \EEq{eq: n5}, except the chiral density $n_5$, as constant in time to derive an analytical solution:
\begin{equation}
   n_5(t) = \left( \frac{2 \alpha }{\pi \hbar} \frac{\EE \cdot \BB}{\Gamma_f} 
   %+ {n_e} \frac{\Gamma_w^\mathrm{eff}}{\Gamma_f} 
   \right) 
   \left(1 - e^{-\Gamma_f t} \right) + n_5^0 e^{-\Gamma_f t},
    \label{eq: n5 analytical solution}
\end{equation}
where $n_5^0$ is the initial chiral number density. On astrophysical timescales ($t \gg \Gamma_f^{-1}$), the system reaches a quasi-equilibrium state:
\begin{equation}
    n_5(t) \approx  
    \frac{2 \alpha }{\pi \hbar} \frac{\EE \cdot \BB}{\Gamma_f} ~.
    %+ {n_e} \frac{\Gamma_w^\mathrm{eff}}{\Gamma_f}.
    \label{eq: n5(t) analytical}
\end{equation}

%  The chiral chemical potential $\mu_5$ can be calculated from $n_5$ as follows \cite{sigl2016}:
%  \begin{equation}
%     \mu_5  = \frac{3 \pi^2 (\hbar c)^3 n_5 }{3 \mu_e^2 + \pi^2 k_b^2 T^2}~, 
%     \label{eq: mu5 from n5}
% \end{equation} 

The chiral chemical potential $\mu_5$ is related to the chiral number density $n_5$ via the standard expression
$n_5 = \mu_e^2 \mu_5 /( \pi^2 (\hbar c)^3 )$, valid in the regime of highly degenerate NS matter where $\mu_e \gg T$.
Substituting the expression for $\EE \cdot \BB$ into \EEq{eq: n5(t) analytical}, we obtain an explicit expression for $k_5$:
\begin{equation}
    k_5\left(\boldsymbol{x},t \right) = \frac{ \left(\Bnabla\times \BB \right) \cdot \BB  }
    {\dfrac{\mu_e^2 }{ 8 \pi \alpha^2 \eta \left(\hbar c\right)} \Gamma_f
    + B^2  } 
    = 
    \frac{ \left(\Bnabla\times \BB \right) \cdot \BB  }
    {\left(\dfrac{2 \mu_e^2 }{m_e^2 c^4}\right) \dfrac{B^2_\mathrm{QED}}{3 \pi}
    + B^2  } ~,
    \label{eq: k5}
\end{equation} 
and $B_\mathrm{QED} \equiv m_e^2 c^3 / (e \,\hbar) = 4.41 \times 10^{13}$\,G is the Schwinger QED critical field. 

\EEq{eq: k5} highlights the two main contributions to the chiral asymmetry: the numerator originates from magnetic helicity and increases when the pointwise current helicity density, $\BB \cdot (\Bnabla\times \BB)$, grows -- that is, when $\Bnabla \times \BB$ becomes increasingly aligned with $\BB$ \cite{blackman2002}. This alignment characterizes the so-called {\it helical} magnetic fields, where toroidal and poloidal components reinforce each other. In contrast, the spin-flip term in the denominator suppresses chiral asymmetry. If the magnetic field configuration possesses a non-zero average magnetic helicity, a small but appreciable chiral asymmetry can be sustained despite the action of spin-flip processes. 
In this regime, where a nonzero chiral asymmetry is sustained and $k_5$ evolves in quasi-equilibrium, one can insert \EEq{eq: k5} into \EEq{Efield}, explicitly factoring out the chiral anomaly from the induction equation. Consequently, 
$k_5$ acts as a catalyst for energy transfer across scales, making the magnetic field evolution depend solely on the field itself.  
In this formulation, the chiral contribution manifests as a nonlinear correction associated with the current component parallel to $\BB$, while the Hall term emerges as a quadratic nonlinearity in $\BB$, governed by the current component perpendicular to the field.

Defining the characteristic wavenumber associated with the current parallel to the magnetic field as
\begin{equation}
    k_{\BB} \equiv \frac{ \left(\Bnabla\times \BB \right) \cdot \BB  }{ B^2} ~,
\end{equation}
we can write
\begin{equation}
    k_5 =  \frac{ k_{\BB} } {1 + \left(\dfrac{2 \mu_e^2 }{ m_e^2 c^4}\right) \dfrac{B^2_\mathrm{QED}}{ 3 \pi B^2} }.
\end{equation}

In the weak-field regime ($B \ll B_\mathrm{QED}$), we find $k_5 \propto k_{\BB} B^2/B_\mathrm{QED}^2 \ll k_{\BB}$. In the strong-field regime ($B > B_\mathrm{QED}$), $k_5$ is ultimately bounded by $k_{\BB}$ ($k_5 \lesssim k_{\BB}$).

To estimate the characteristic magnetic field strength at which saturation sets in, we equate the two terms in the denominator of \EEq{eq: k5}. This condition is satisfied when:
\begin{equation}
B_\mathrm{sat} \approx \sqrt{\frac{2}{3 \pi}} \frac{\mu_e }{m_e c^2} B_\mathrm{QED}.
    \label{eq: Bmax}
\end{equation}
The saturation field $B_\mathrm{sat}$ scales linearly with the ratio $\mu_e/(m_e c^2)$, which increases with density. Under typical NS conditions, this ratio ranges from 10 in the outer crust to 200 in the inner crust, yielding $B_\mathrm{sat} \sim 10^{14}$\,G near the surface and up to $\sim 5 \times 10^{15}$\,G in deeper layers---consistent with inferred magnetar field strengths.

It is insightful to explicitly write the conservation equations for electromagnetic energy and the additional electron energy from chiral imbalance (see \cite{kaplan2017} for a thorough discussion on energy conservation):
\begin{eqnarray}
\frac{\partial \varepsilon_{em}}{\partial t}  &=&  
 - \sigma_e {\EE^2} - \frac{\alpha \mu_5}{\pi \hbar} \EE \cdot \BB
 - \frac{c}{4 \pi}  \Bnabla \cdot \left( \EE  \times \BB  \right),
\nonumber \\
  \frac{\partial\varepsilon_5}{\partial t}  &=& 
     -  \frac{1}{2} \mu_5 n_5 \Gamma_f + 
    \frac{\alpha \mu_5}{\pi \hbar} \EE \cdot \BB   ~.
    \label{eq: energy density}
\end{eqnarray}
Here, $\varepsilon_{em} = B^2 / 8\pi$ and $\varepsilon_5 = \mu_5 n_5 / 2$, with each term on the right-hand side plays a distinct role. The first term in both equations represents a sink: Joule dissipation for magnetic energy and spin-flip processes for chiral energy. The final term in the magnetic energy equation corresponds to the Poynting flux, which becomes a surface term upon volume integration. The $\EE \cdot \BB$ terms describe the conservative exchange of energy between the magnetic field and chirality.

Integrating over the stellar volume, the total energy balance reads: 
\begin{equation}
\frac{d}{dt} \left(E_{em} + E_5\right) + S_\mathrm{tot} + Q_\mathrm{tot} + {\Gamma_5}_\mathrm{tot}  = 0,
\label{eq: total energy}
\end{equation}
where $E_{em}$ and $E_5$ are the volume integrals of $\varepsilon_{em}$ and $\varepsilon_5$, respectively. $S_\mathrm{tot} = \dfrac{c}{4 \pi} \oint dS \cdot \left( \EE \times \BB\right)$ is the total Poynting flux, ${\Gamma_5}_\mathrm{tot} = \frac{1}{2} \int \mu_5 n_5 \Gamma_f \, dV $ is the total spin-flip dissipation rate, and $Q_{\mathrm{tot}}$ is the total Joule dissipation, given by:
\begin{eqnarray}
Q_\mathrm{tot} = \int \sigma_e \EE^2 dV .  
    \label{eq: Qtot}
\end{eqnarray}
In the absence of the Hall drift term, $ c\, \EE = \eta \left( \Bnabla \times \BB - k_5  \BB \right)$, and the chiral correction can offset magnetic dissipation if $\nabla \times \BB \approx  k_5 \BB$. 

% We can substitute the expression for $k_5$ from \EEq{eq: k5} into equation (\ref{Efield}), and after some cancellations and reorganization, we obtain
%\begin{equation}
%    \EE \cdot \BB =   \left[ \frac{1}{1 + 
%    \dfrac{B^2}{B^2_\mathrm{sat}} }\right] \frac{\eta}{c} \, \BB \cdot (\nabla \times \BB) .
%\end{equation}

To assess the impact of the CME on magnetic field evolution in NSs, we performed magneto-thermal simulations using an extended 3D finite-volume version of the \MATINS\, code \cite{dehman2023,dehman2023b,ascenzi2024}. This version solves the coupled induction and heat diffusion equations, consistently incorporating chiral and spin-flip terms arising from the chiral anomaly. 
% General relativistic corrections are included in the simulations but are omitted from the equations shown here for clarity.
To focus on the NS crust, we apply potential-field boundary conditions (current-free magnetosphere) at an outer numerical boundary located at $\rho = 10^{10}\,\mathrm{g\,cm}^{-3}$, near the transition between the liquid envelope and the solid crust. At the crust–core interface, we impose perfect-conductor boundary conditions. The temperature-dependent electrical conductivity is computed at each point of the star using the codes from the IOFFE repository\footnote{\url{http://www.ioffe.ru/astro/conduct/}} 
\cite{potekhin2015}. In \MATINS, the NS background model can be constructed using various zero-temperature equations of state (EOS) from the CompOSE online database\footnote{\url{https:/compose.obspm.fr/}}. 
For this study, we adopt the BSk24 EOS~\cite{pearson2018}, assuming a canonical NS with mass $M = 1.4\,M_{\odot}$, radius $R = 12.4$\,km, and crustal thickness of 0.86\,km.

Magnetic helicity plays a central role in magnetic field dynamics~\cite{brandenburg2005}, enabling inverse cascades that transfer energy from small to large scales. In NS crusts, such cascades are typically driven by the Hall effect on Hall timescales~\cite{brandenburg2020,dehman2025}. However, Dehman \& Brandenburg~\cite{dehman2025} showed that, although present, this process fails to significantly amplify the large-scale dipolar field due to the crust’s extreme radial-to-angular aspect ratio.
To isolate and better assess the role of chirality, we deliberately disable the nonlinear Hall term and focus solely on the impact of the CME.

%%%%%%%%%%%%%%%%%%%%%%%%%%%%%%%%%%%%%%%%%%%%%%%%%%%%%%
% Initial magnetic field 
%%%%%%%%%%%%%%%%%%%%%%%%%%%%%%%%%%%%%%%%%%%%%%%%%%%%%%
As detailed in Section~\ref{app: helical field} of the Supplemental Material, we initialize our simulations with a turbulent, partially helical\footnote{
A single-mode helical field is force-free, with the current everywhere parallel to the magnetic field 
\big($(\Bnabla \times \BB) \parallel \BB$\big), 
leading to a zero Lorentz force. However, combining modes with different wavenumbers generally does not produce a force-free condition due to mode interactions.} magnetic field (see Appendix A of Reisenegger~\cite{reisenegger2009}) with $B \gg B_{\rm sat}$, concentrated at very small scales and with an average strength of $\sim 10^{16}$\,G~\cite{reboul2021,masada2022}.
The field is initially divided between poloidal and toroidal components, with a dominant toroidal contribution, consistent with the small-scale energy spectra predicted by proto-NS dynamo simulations~\cite{reboul2021}.
We limit the dipolar component to a few $10^{12}$\,G, much weaker than the large-scale fields inferred observationally for magnetars.
The total magnetic energy, of order a few $10^{49}$\,erg, remains small compared to the star’s gravitational binding energy ($\sim 10^{53}$\,erg) or the rotational energy of a millisecond pulsar ($\sim 10^{51-52}$\,erg).

The magnetic energy spectrum peaks at $\ell_0 \approx 50$, extends up to $\ell_{\mathrm{max}} = 70$, and follows a $\ell^4$ slope. We consider three initial setups: (i) Full Slope (Run~F), preserving the full $\ell^4$ spectrum and naturally concentrating energy near $\ell_0$; (ii) Damped (Run~D), where the magnetic energy is attenuated by roughly two orders of magnitude for $\ell = 1\ldots20$, concentrating the remaining energy between $\ell = 21$ and $\ell_0$; and (iii) Damped–No CME (Run~DO), a replica of Run~D without CME. 
The radial wavenumber is set to $k_r\approx 400\,\mathrm{km}^{-1}$ (Figure~\ref{fig:k_peak}), balancing the fastest-growing CME modes -- characterized by chiral wavelengths of a few meters -- against Ohmic dissipation. To resolve these small scales, we adopt $N_r = 200$ radial points, reflecting the CME’s strong sensitivity to microphysical properties (e.g., $\eta$ and $\mu_e$), which vary primarily with radius. Angular directions are discretized using a cubed-sphere grid~\cite{dehman2023}, with $N_\xi = N_\eta = 47$ points per patch across six patches, yielding $N_\theta = 94$ and $N_\phi = 188$ grid points, and resolving angular structures down to a few hundred meters. 
Starting from these initial conditions, we run three global NS crust simulations, evolving each over the first few hundred years with a timestep of days---consistent with the ages of the youngest magnetars, such as Swift J1818.0–1607 ($\sim 200$ yr;~\cite{esposito2020}).

\begin{figure*}[ht!]
    \centering
\includegraphics[width=0.45\textwidth]{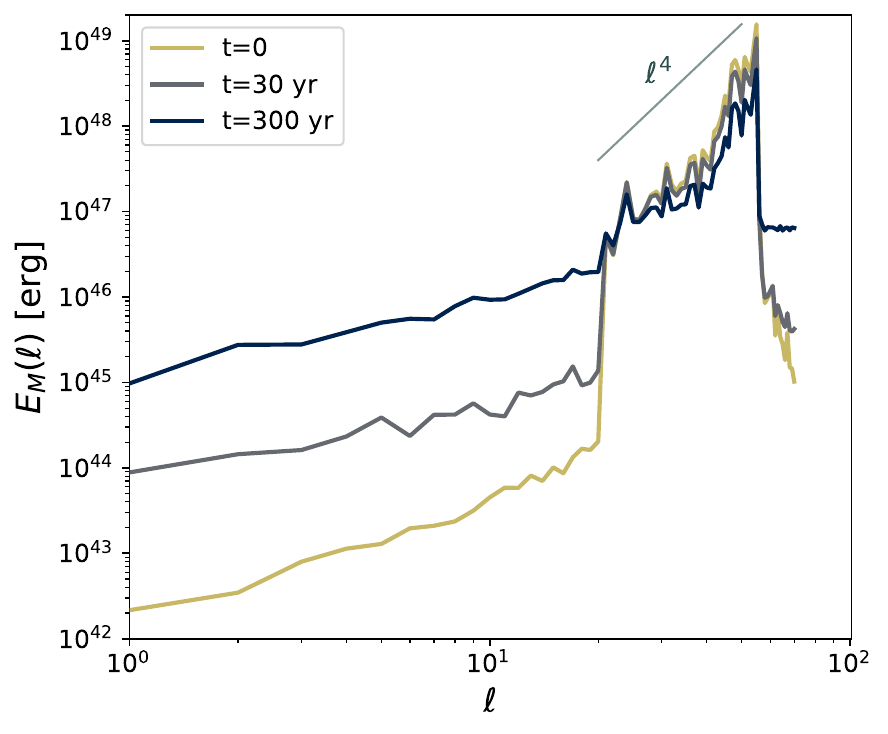}
\includegraphics[width=0.45\textwidth]{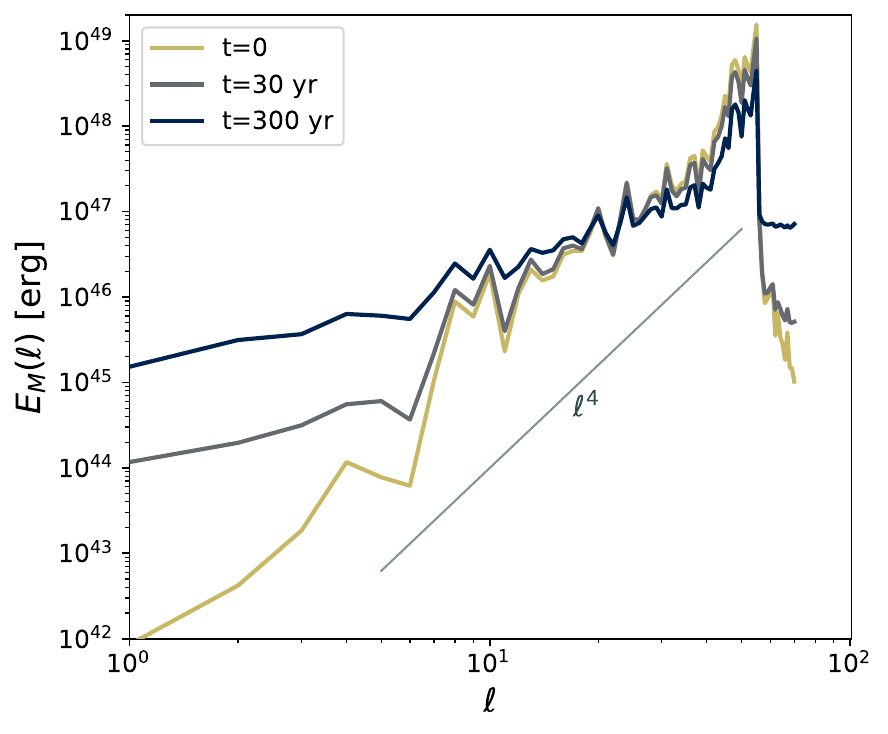}
    \caption{Magnetic energy spectra for Run~D (left) and Run~F (right) as a function of multipole degree $\ell$. Spectra are shown at $t=0$ (yellow), $t=30$\,yr (dark gray), and $t=300$\,yr (dark blue), with a reference $\ell^4$ slope overplotted in light gray.
   } 
    \label{fig: spectral energy}
\end{figure*}
%%%%%%%%%%%%%%%%%%%%%%%%%%%%%%%%%%%%%%%%%%%%%%%%%%%%%%
% Results description
%%%%%%%%%%%%%%%%%%%%%%%%%%%%%%%%%%%%%%%%%%%%%%%%%%%%%%
In the absence of spin-flip processes (an unphysical case), the axial charge $Q_5$ would grow to match the initial magnetic helicity $\left(\frac{\alpha}{\pi\,\hbar\,c}\,\chi_m\right)$, conserving total helicity. However, spin-flip scattering restores chiral balance, suppressing $Q_5$ by nearly 20 orders of magnitude, leaving a tiny residual -- sufficient to drive chiral-induced magnetic field evolution. 
See Section~\ref{app: detailed analysis} in the Supplemental Material for details.

To assess the impact of this residual asymmetry, Figure~\ref{fig: spectral energy} shows the total magnetic energy spectra for Run~D (left) and Run~F (right) at $t=0$, $t=30$ yr, and $t=300$ yr. A significant energy transfer---about two orders of magnitude---occurs toward multipoles with initially weak fields ($\ell \leq 20$), as the emergence of a nonzero $k_5$ enables the CME to redistribute magnetic energy across spatial scales, boosting weaker regions at the expense of stronger ones. A particularly striking feature is the strong amplification of the dipolar component ($\ell=1$), marking the natural formation of the largest and hardest-to-form scale. In Run~D, the initial slope, nearly linear in $\ell$ for $\ell \leq 20$, is preserved but amplified. Conversely, in Run~F, the slope at large scales gradually evolves toward a similar linear dependence on $\ell$. Ultimately, both runs yield comparable magnetic energy spectra, suggesting that each mode saturates at a distinct characteristic amplitude.
In both cases, the CME preserves the spectral peak at $\ell_0$, which -- being at small scales -- is more susceptible to Ohmic dissipation over tens of kiloyears. By contrast, an inverse cascade typically shifts the peak toward lower $\ell$, enhancing the longevity of the large-scale field~\cite{brandenburg2020}.

While this process may resemble an inverse cascade, its underlying mechanism is fundamentally different. 
In the presence of a helical magnetic field (as in our configuration) and an active Hall term, an inverse cascade can occur, though its impact is typically much weaker~\cite{dehman2025}. In such cases, nonlinear interactions---specifically those involving the component of the current density perpendicular to the magnetic field---transfer energy and helicity to larger scales through mode couplings that satisfy $\boldsymbol{k} = \boldsymbol{p} + \boldsymbol{q}$ with $|\boldsymbol{k}| \leq \max(|\boldsymbol{p}|, |\boldsymbol{q}|)$, as demonstrated by Frisch~\cite{frisch1975}. By contrast, the CME, which involves the component of the electric current parallel to the magnetic field, operates simultaneously and independently on all multipoles (see Section~\ref{app: pol-tor induction eq} of the Supplemental Material), enabling the redistribution of magnetic energy from small to larger scales in our simulations.

\begin{figure}[ht!]
    \centering
\includegraphics[width=\linewidth]{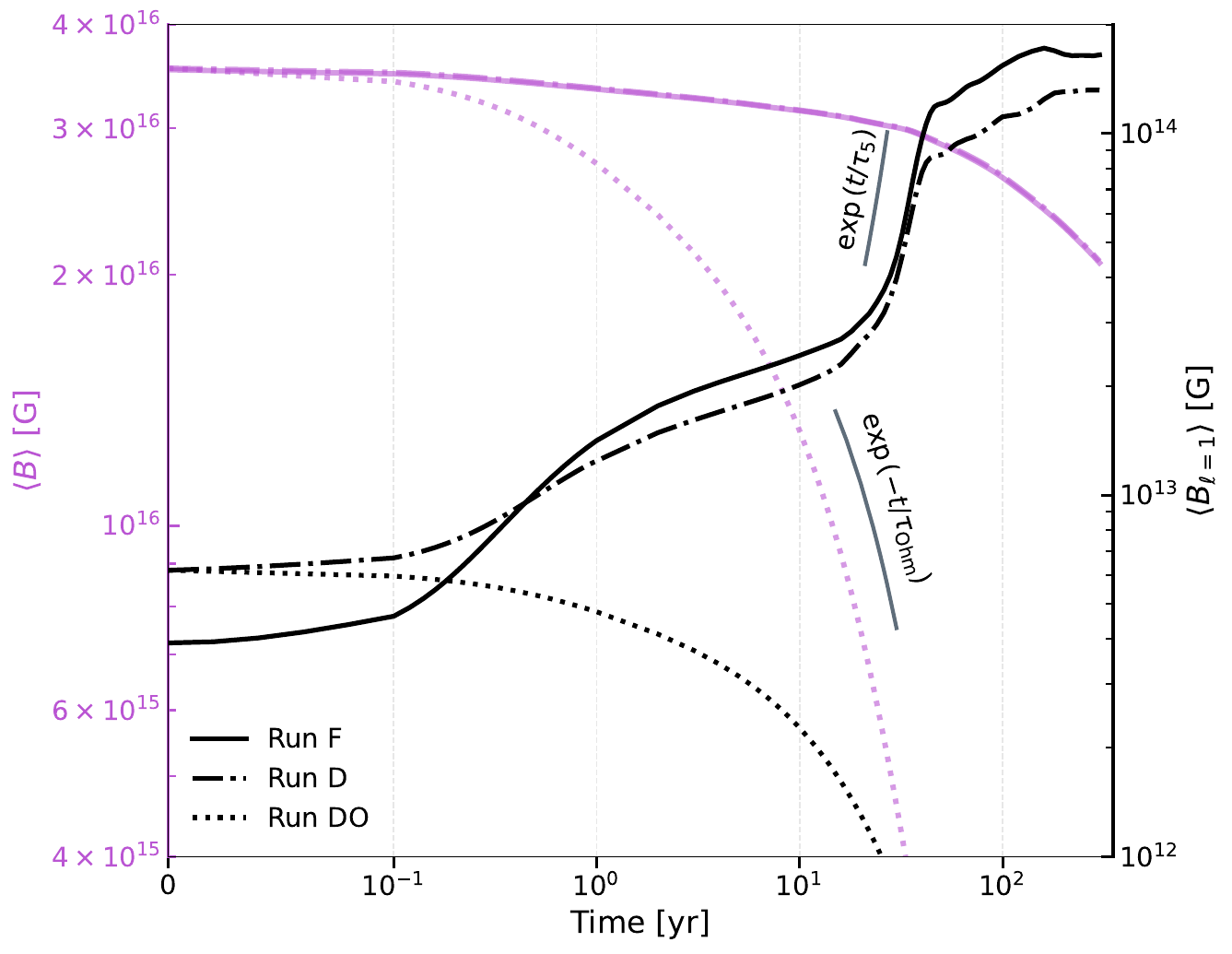}
    \caption{Time evolution of the average magnetic field (mauve, left axis; scaling from $4 \times 10^{15}$\,G to $4 \times 10^{16}$\,G) and dipolar field (black, right axis; scaling from $10^{12}$\,G to $2 \times 10^{14}$\,G). Solid lines correspond to Run~F, dash-dotted lines to Run~D, and dotted lines to Run~DO. Gray lines show fits to the growth and decay phases, with $\tau_\mathrm{Ohm} \equiv 1/\eta k^2  \approx 20 \dots 25$ yr and $\tau_5 \equiv 1/\eta k k_5  \approx 5 \dots 10$ yr. }
    \label{fig: pol tor energy dipole}
\end{figure}

Figure~\ref{fig: pol tor energy dipole} illustrates the decay of the average magnetic field (mauve, left axis)
alongside the growth of the dipolar field (black, right axis). Note the different scales on the left and right vertical axes. Run~F, Run~D, and Run~DO are represented with solid, dash-dotted, and dotted lines, respectively. 

As expected, the average magnetic field decreases over time in all three runs. 
For reference, we show with dots Run~DO including only Ohmic dissipation. We observe a rapid decay of the initial small-scale magnetic field following $\propto \exp\left( -t/\tau_\mathrm{Ohm} \right),$ with $\tau_\mathrm{Ohm} \equiv 1/\eta k^2 \approx 20\ldots25$ yr. The initially weak dipolar field also undergoes decay, though at a slower rate due to its smaller wavenumber $k$.

In contrast, the CME is active in Runs~F and D, transferring a portion of the energy from the small-scale initial 
magnetic field to the dipolar component, which consequently grows over time. It is worth noting that overall dissipation is significantly slower than in Run~DO, as the self-regulating chiral current effectively counteracts much of the original electric current along the magnetic field lines, thereby reducing the net Joule dissipation rate (\Eq{eq: Qtot}).

In Runs~F and D, we identify three distinct stages in the growth of the dipolar components. During the initial few months ($t \leq 0.1$\,yr), little noticeable difference appears, as the chiral asymmetry is still developing from the initial turbulent average field. This is followed by a gradual growth phase lasting 20 to 30 years, and then by a second, exponential growth phase spanning several decades. We interpret this latter phase as a manifestation of the chiral magnetic instability (CMI), with the growth approximately following $\propto \exp(t/\tau_5)$, with $\tau_5 \equiv 1/\eta k k_5 \approx 5 \ldots 10$ yr. As expected, the growth saturates at $ B \gtrsim 10^{14}$\,G after $\sim 100$ yr, consistent with magnetar observations. For a detailed analysis of the magnetic field evolution in Run~F, see Section~\ref{app: detailed analysis} of the
Supplemental Material.

% %%%%%%%%%%%%%%%%%%%%%%%%%%%%%%
% Conclusion
% %%%%%%%%%%%%%%%%%%%%%%%%%%%%%%

Our research reveals that magnetic helicity, a classical attribute, triggers a subtle chiral asymmetry---a quantum anomaly---in macroscopic astrophysical bodies like NSs. This faint chirality significantly shapes magnetic field evolution over centuries, overcoming robust spin-flip suppression. 
The CME restructures the magnetic field spectrum by redistributing energy across scales, converging toward a nearly linear slope in multipole degree $\ell$, particularly for $\ell < 10$, which governs large-scale astrophysical observables. This linearity arises because each mode saturates at its characteristic amplitude.

An initial intense ($10^{16}$\,G) very small-scale field naturally evolves into a mixed poloidal/toroidal dipolar field, amplified to $10^{14}$\,G---consistent with observed magnetar strengths. 
Although small-scale fields initially persist after this short stage (lasting 100 years), they are more prone to Ohmic decay and will dissipated in a few thousand years. This strong small-scale field may potentially drive the high X-ray luminosities ($L_X \gtrsim 10^{35}$\,erg/s) and burst activity, due to energy stored at small scales~\cite{dehman2020}. Conversely, the large-scale dipolar field decays more gradually, persisting as a long-lived feature. This interaction between magnetic helicity and the chiral anomaly establishes a novel framework for understanding magnetar field evolution.

A key insight from our study is the geometry of the resulting magnetic field. Previous research
~\cite{Geppert2012,Vigano2013,Gourg2019,dehman2020,deGrandis2021,dehman2023,dehman2023b} on long-term magneto-thermal evolution indicate that crust-confined fields with prominent large-scale structures align best with observational data. 
However, generating such configurations is difficult, as conventional proto-NS dynamo processes create intense small-scale turbulence but struggle to produce crust-confined large-scale dipolar fields. Our work introduces a mechanism that naturally achieves this geometry. We have presented the first self-consistent, long-term simulations of the CME-driven magnetic evolution, establishing this mechanism as a key ingredient for magnetar dynamics. 

Enhancing numerical resolution at small scales is anticipated to amplify the observed effects further, and we plan to refine this in future simulations. This study specifically isolates the CME under strong magnetic fields by deliberately excluding the nonlinear Hall term, enabling a clear analysis of the CME role. A future investigation will explore the impact of the Hall effect on the chiral anomaly. Preliminary results suggest that the Hall effect has a limited influence on the early-time (100 years) evolution, the focus of this paper, as it generally operates over longer timescales.

\begin{acknowledgments}
We thank anonymous referee B for a careful report that improved the manuscript. 
We gratefully acknowledge stimulating discussions with Axel Brandenburg, Sanjay Reddy, Andrei Beloborodov, Chris Thompson and Valentin Skoutnev and other participants at the \emph{``IReNA‐INT Joint Workshop on Thermal and Magnetic Evolution of Neutron Stars'}' (University of Washington), \emph{``Breaking New Ground in Supernova Physics 2025: Crossroads of Turbulence, Chiral Dynamics, and Machine Learning''} (Fukuoka University), and \emph{``Extreme Physics of Neutron Star Interiors''} (Princeton University). We thank the organizers for their hospitality and for inviting CD as a speaker. CD is supported by the Ministerio de Ciencia, Innovación y Universidades (JDC2023-052227-I), co-funded by AEI (MCIN/AEI/10.13039/501100011033), the FSE+, and the Universidad de Alicante. CD and JP acknowledge support from the Conselleria d'Educació, Cultura, Universitats i Ocupació de la Generalitat Valenciana, through grant CIPROM/2022/13, and from the AEI grant PID2021-127495NB-I00 funded by MCIN/AEI/10.13039/501100011033.
\end{acknowledgments}

\bibliography{apssamp}
\nocite{apsrev42Control}
\newpage
\onecolumngrid
\begin{center}
\textbf{\large Supplemental Material}
\end{center}
\makeatletter

In the subsequent sections, we initially break down the induction equation (omitting the Hall term) into its poloidal and toroidal components to show how the chiral current links the different modes (Section~\ref{app: pol-tor induction eq}). 
Although Section~\ref{app: pol-tor induction eq} 
presents a simplified toy model, treating the chiral wavenumber $k_5$ and magnetic diffusivity $\eta$ as constants 
for a clear, quasi-analytical illustration of the proposed mechanism, the simulations themselves were performed using a full 3D code, without these simplifying assumptions.
We then describe the setup of the initial magnetic field used in this study (Section~\ref{app: helical field}).

\section{Simplified discussion of the CME in the induction equation}
\label{app: pol-tor induction eq}
The magnetic field can be decomposed into poloidal ($\BB_p$) and toroidal ($\BB_t$) components as follows ~\cite{chandrasekhar1957}:
\begin{equation}
\BB = \BB_p + \BB_t,
\end{equation}
\begin{eqnarray}
  &&  \BB_t = - \boldsymbol{r} \times \Bnabla \Psi,  \quad \quad \quad   \AAA_t = - \rr \times \Bnabla \Phi, \nonumber \\
  &&   \BB_p = \Bnabla \times \AAA_t = - \boldsymbol{r} \Delta \Phi + \Bnabla \frac{\partial }{\partial r} \left(r \Phi \right),
\label{eq: pol tor relations}
\end{eqnarray}
where $\AAA_t$ is the toroidal vector potential. The two scalar functions $\Phi(\rr,t)$ and $\Psi(\rr,t)$ uniquely define the poloidal and toroidal components, respectively.

The induction equation (\Eq{eq: faraday law}) can also be decomposed into poloidal and toroidal parts. Neglecting the Hall term and retaining only the Ohmic and chiral contributions, one can follow the same methodology as in Geppert \& Wiebicke~\cite{geppert1991} to obtain
\begin{eqnarray}
\frac{\partial \Phi}{ \partial t} &=&  \eta \left( \Delta \Phi +  k_5 \, \Psi \right) ,  \nonumber \\
\frac{\partial \Psi}{ \partial t}  &=& \eta \left( \Delta \Psi - k_5 \Delta \Phi \right) .
   \label{eq: pol tor ind eq 2}
\end{eqnarray}

The additional term proportional to $k_5$ accounts for the CME.
For clarity, we have omitted the spatial (radial and angular) dependence of the magnetic diffusivity $\eta$ and the chiral wavenumber $k_5$, and assumed they are constants.

Next, we expand the scalar functions in spherical harmonics \cite{KR80}: 
\begin{eqnarray}
    \Phi(t,r,\theta,\phi) &=& \frac{1}{r}\sum_{\ell m} \Phi_{\ell m}(r,t) Y_{\ell m}(\theta,\phi),  \nonumber\\
    \Psi(t,r,\theta,\phi)  &=& \frac{1}{r}\sum_{\ell m} \Psi_{\ell m}(r,t) Y_{\ell m}(\theta,\phi), 
       \label{eq: phi and Psi scalar functions}
  \end{eqnarray}
where $\ell = 1, 2, \ldots$ denotes the multipole degree and $m = -\ell, \ldots, \ell$ the azimuthal order. 
\EEq{eq: pol tor ind eq 2} yields a set of equations that couples the toroidal and poloidal components of each multipole, taking the following form:
\begin{eqnarray}
    \frac{\partial \Phi_{\ell m}}{\partial t} &=& \eta ~\Delta \Phi_{\ell m}  +  \eta \, k_5 \, \Psi_{lm} , \nonumber\\
    \frac{\partial  \Psi_{\ell m}}{\partial t} &=& \eta ~\Delta \Psi_{\ell m} - \eta \, k_5 \, \Delta \Phi_{\ell m} ,
\end{eqnarray}
where 
\begin{equation}
    \Delta \equiv \left( \dfrac{\partial^2 }{\partial r^2} - \dfrac{\ell \left(\ell +1 \right)}{r^2}  \right)~.
\end{equation} 
Note that incorporating the spatial dependence of $\eta$ and $k_5$ would introduce additional coupling terms in the equations.
In this form, the role of $k_5$ becomes evident.
First, it couples independently to all multipoles $(\ell, m)$ of the toroidal and poloidal components.
Second, the evolution equations show that an initial poloidal (or toroidal) field acts as a source in the toroidal (or poloidal) evolution equation, naturally generating its complementary counterpart and driving the system toward an approximate equipartition of magnetic energy between the two. However, this coupling is inherently asymmetric: the poloidal field couples directly to the toroidal magnetic component, while the toroidal field couples to the toroidal current (i.e., the curl of the poloidal component). 

To simplify the discussion, we introduce additional notation. The Laplacian operator acts on each mode as
\begin{equation}
    \Delta \rightarrow -(k_r^2+k_{\mathrm ang}^2),
\end{equation}
where $k_r$ characterizes the radial dependence (i.e., $\partial / \partial r \rightarrow i k_r$), and the angular wavenumber is $k_{\mathrm{ang}} = \sqrt{\ell(\ell + 1)}/r$, reducing the equations to:
 \begin{eqnarray}
    \frac{\partial \Phi_{\ell m}}{\partial t} &=& - \eta ~ 
    (k^2 \Phi_{\ell m}  -  \, k_5  \Psi_{lm}) ,
\nonumber \\
    \frac{\partial  \Psi_{\ell m}}{\partial t} &=& - \eta ~
    (k^2 \Psi_{\ell m} -  k_5 ~ k^2 \Phi_{\ell m}) .
\end{eqnarray}
Consider a field satisfying the condition
$\Psi_{\ell m} = k \, \Phi_{\ell m}$. The equations then become
 \begin{eqnarray}
    \frac{\partial \Phi_{\ell m}}{\partial t} &=& - \eta ~ 
    k (k - k_5)  \Phi_{lm} ,
\nonumber \\
    \frac{\partial  \Psi_{\ell m}}{\partial t} &=& - \eta ~
    k (k -  k_5) \Psi_{\ell m} ,
\end{eqnarray}
which admit exponentially growing or damped modes, depending on the sign of $k - k_5$.
For $k > k_5$, the instability is suppressed by Ohmic dissipation (if $k_5$ is negative, dissipation is even enhanced).
In contrast, for $k < k_5$, the CMI develops, and the fastest-growing chiral modes ($k \approx k_5 / 2$) dominate.
These fastest-growing chiral modes are associated with small-scale magnetic structures (typically ranging from centimeters to meters), characterized by the CME wavelength:
\begin{equation}
\lambda \equiv \frac{\pi}{k_5} = \frac{\pi \hbar c}{4 \alpha \mu_5}~.
\label{eq: lambda_5 app}
\end{equation}

\section{Initial magnetic field structure and optimal CME mode}
\label{app: helical field}
%%%%%%%%%%%%%%%%%%%%%%%%%%%%%%%%%%%%%%%%%%%%%%%%%%%%%%%%%%%%%
% Initial field construction -- before describing the figure 
%%%%%%%%%%%%%%%%%%%%%%%%%%%%%%%%%%%%%%%%%%%%%%%%%%%%%%%%%%%%%
 \begin{figure}[ht!]
    \centering
\includegraphics[width=0.8\linewidth]{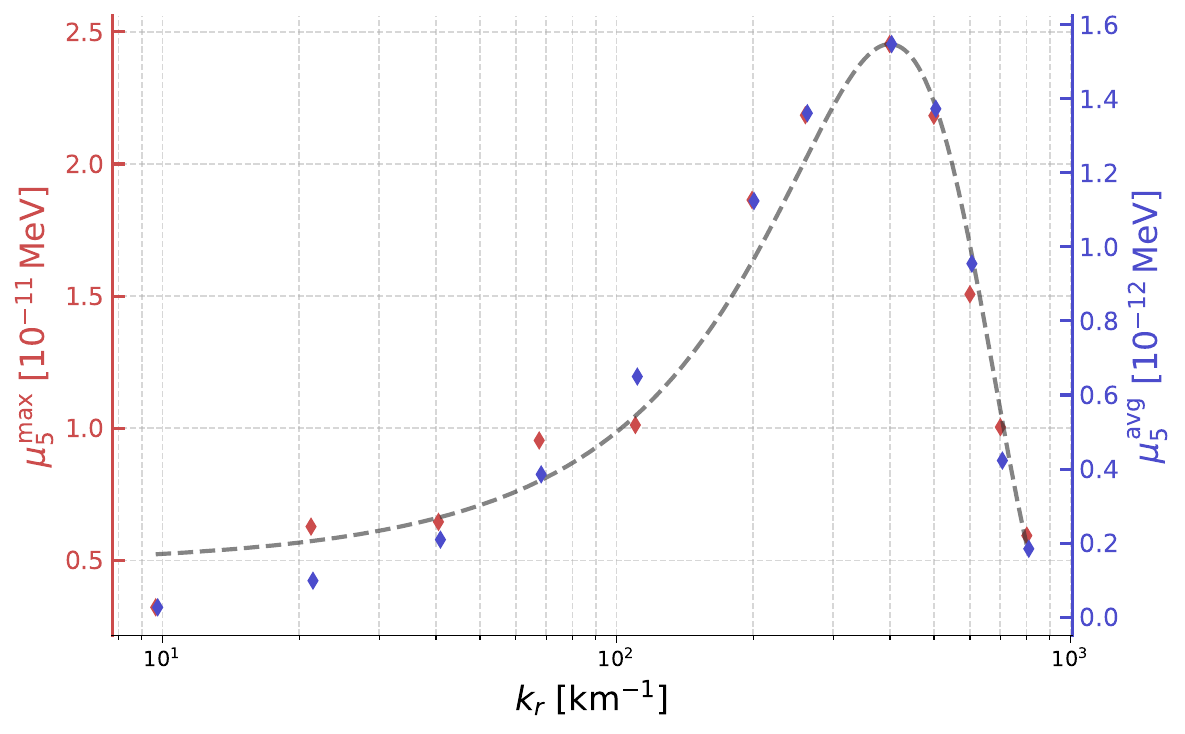}
    \caption{\label{fig:k_peak}  
 Maximum ($\mu_5^{\mathrm{max}}$ in units of $10^{-11}$ MeV; red diamonds) and average ($\mu_5^{\mathrm{avg}}$ in units of $10^{-12}$ MeV; blue diamonds) chiral chemical potentials as functions of the radial wavenumber $k_r$, computed at the initial time for an average magnetic field of a few $10^{16}$\, and a temperature of $T=10^9$\,K. The peak occurs at $k_0 \approx 400\ \mathrm{km}^{-1}$, corresponding to a wavelength $\lambda \approx 14\ \mathrm{m}$. The gray dashed curve shows an asymmetric Gaussian fit, $\mu_5^{\mathrm{max}}(k_r) = \mu_5^0 \exp\left[ - (k_r - k_0)^2 / (2\sigma_\pm^2) \right]$, with different widths $\sigma_-$ and $\sigma_+$ on either side of the peak. Here, $\mu_5^0 \equiv \mu_5^{\mathrm{max}}(k_0) \approx 2.5 \times 10^{-11}\ \mathrm{MeV}$.} 
\end{figure}

All the microphysical properties relevant for the CME in the NS interior (e.g., magnetic diffusivity $\eta$, electron chemical potential $\mu_e$), vary rapidly in the radial direction, and to a lesser extent in angular directions due to possible temperature anisotropies. Since the crust is very thin compared to the star's perimeter, radial gradients are expected to dominate the dynamics. This highlights the need for higher radial resolutions
in the simulations.  Thus, one can expect 
$k_r \gg k_{\mathrm ang}$, and we can approximate $ k^2 \approx k_r^2$.

Guided by this insight, we construct the initial magnetic field in the NS crust using \MATINS\, by prescribing the angular and radial profiles of the scalar functions $\Phi(\rr, t{=}0)$ and $\Psi(\rr, t{=}0)$ (see \Eq{eq: phi and Psi scalar functions}). 

The radial function $\Phi_{\ell m}$ takes the following form~\cite{aguilera2008}: 
\begin{equation} 
\Phi_{\ell m}(r) = \Phi^0_{\ell m} \, k_r r \, (a + \tan(k_r R) \, b), 
\label{eq: funa function} 
\end{equation} 
where $a$ and $b$ are chosen to satisfy the inner and outer boundary conditions \cite{aguilera2008}, and $\Phi^0_{\ell m}$ are normalization factors evaluated at a reference radius, just beneath the stellar surface $R_\star$.
These weights are chosen to concentrate magnetic energy at small angular scales, following a spectral slope proportional to $\ell^4$ in the sub-inertial range ($\ell < \ell_0$), where $\ell_0$ denotes the spectral peak \cite{DC03}.

We encode magnetic helicity by directly relating the toroidal scalar function via:
\begin{equation}
\Psi_{\ell m}(r) =  \alpha_{\ell m} \, \Phi_{\ell m}(r), 
\label{eq: Phi = k Phi}
\end{equation}
where, being conservative, we set $\alpha_{\ell m}= \sqrt{\ell(\ell+1)}/R~$, where $R$ is the radius at the surface of our computational domain. For a maximally helical field, one should choose $\alpha_{\ell m} = k$.

Employing curl operators adapted to cubed-sphere coordinates~\cite{dehman2023}, we derive the magnetic field components from these scalar functions using \EEq{eq: pol tor relations}. This method guarantees an initial magnetic field that is divergence-free (to machine precision), free of axis singularities, and intrinsically helical---a key property for investigating CME-driven magnetic evolution.

%%%%%%%%%%%%%%%%%%%%%%%%%%%%%%%%%%%%%%%%%%%%%
% Identifying the optimal radial wavenumber
%%%%%%%%%%%%%%%%%%%%%%%%%%%%%%%%%%%%%%%%%%%%%

The radial direction in the crust is resolved with 200 grid points, capturing structures down to a few meters. Angular directions are discretized using a cubed-sphere grid~\cite{dehman2023}, with $N_\xi = N_\eta = 47$ points per patch across six patches, yielding $N_\theta = 94$ and $N_\phi = 188$ grid points, and resolving multipoles up to $\ell_\mathrm{max} \sim 70$, corresponding to angular scales of a few hundred meters. With this setup, we vary the radial wavenumber $k_r$ while keeping the initial average magnetic field and temperature fixed. For each configuration, we compute the predicted average ($\mu_5^{\mathrm{avg}}$) and maximum ($\mu_5^{\mathrm{max}}$) values of the initial chiral chemical potential $\mu_5(\xx)$.

Figure~\ref{fig:k_peak} summarizes the results: maximum (red diamonds) and average (blue diamonds) values of $\mu_5$ as functions of $k_r$, evaluated for a mean magnetic field of a few $10^{16}$\,G concentrated at small scales and a temperature of $10^9$\,K---typical of young magnetars. The optimal radial wavenumber is found at $k_0 \approx 400\, \mathrm{km}^{-1}$, corresponding to a wavelength $\lambda_0 \approx 14$ m, and we ensure that this scale is resolved by our radial grid. Notably, the values of $\mu_5$ are many orders of magnitude smaller than the electron chemical potential ($\mu_e \approx 10\ldots100$ MeV), yet they lead to significant changes in the field evolution.

% %%%%%%%%%%%%%%%%%%%%%%%%%%%%%%
% Detailed analysis of Run~F
% %%%%%%%%%%%%%%%%%%%%%%%%%%%%%%
\section{DETAILED ANALYSIS OF MAGNETIC FIELD EVOLUTION}
\label{app: detailed analysis}

In this section, we present an extended analysis of Run~F. We examine the growth of the dipolar magnetic field, the decay of the mean magnetic field, and the energy transfer between poloidal and toroidal components. We also assess the conservation of total helicity and energy, as outlined in our theoretical framework.

\begin{figure*}[ht!]
    \centering
   \includegraphics[width=\linewidth]{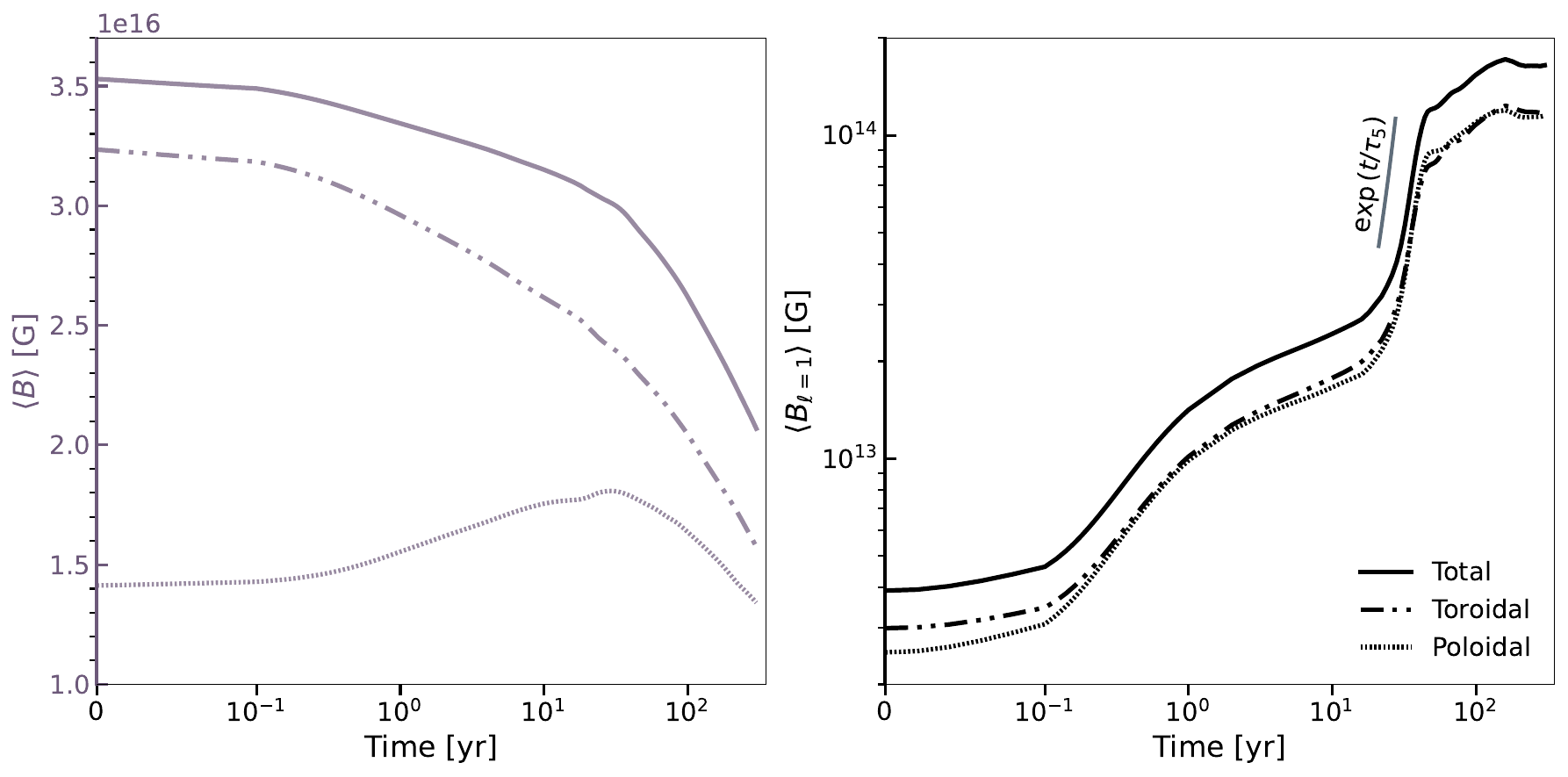}
    \caption{Time evolution of the mean magnetic field (left panel) and the dipolar magnetic field (right panel) for Run~F. Solid lines show the total field, dash–double-dotted lines the toroidal component, and dotted lines the poloidal component. 
    }
    \label{fig: pol tor run F}
\end{figure*}

Figure~\ref{fig: pol tor run F} shows the time evolution of the average magnetic field (left panel) and the dipolar magnetic field (right panel) for Run~F. The toroidal (dash-double-dot) and poloidal (dotted) contributions are also shown. In contrast to Figure~\ref{fig: pol tor energy dipole} in the main text, this figure highlights the interplay between the toroidal and poloidal components. Three distinct stages are observed in the decay of the total magnetic field and the growth of the dipolar field:
i) Early stage ($t \lesssim 0.1$ yr): Both the total and dipolar magnetic fields remain nearly constant. This reflects the initial buildup of the CME, during which the chiral asymmetry is still developing and its dynamical effects are negligible. 
ii) Intermediate stage (up to $\sim 30$ yr): A modest decline in the total magnetic field begins, primarily due to the dissipation of the dominant toroidal component. During this stage, the CME becomes active, transferring energy from the toroidal field to the initially subdominant poloidal field. At the same time, we observe in the right panel the simultaneous growth of both components of the dipolar field. Energy is continuously exchanged between them, maintaining an approximate equipartition.
iii) Late stage (beyond $\sim 30$ yr): At this point, the growth of the poloidal component in the total field halts. Both components have reached similar strengths and begin to dissipate at comparable rates. Meanwhile, the dipolar field exhibits exponential growth, $\propto \exp(t/\tau_5)$, with $\tau_5 \approx 5 \ldots 10$ years---signaling the onset of the CMI. During this process, both components of the dipolar field reach $10^{14}$ G. After about a hundred years, the growth saturates.

\begin{figure*}[ht!]
    \centering
   \includegraphics[width=\linewidth]{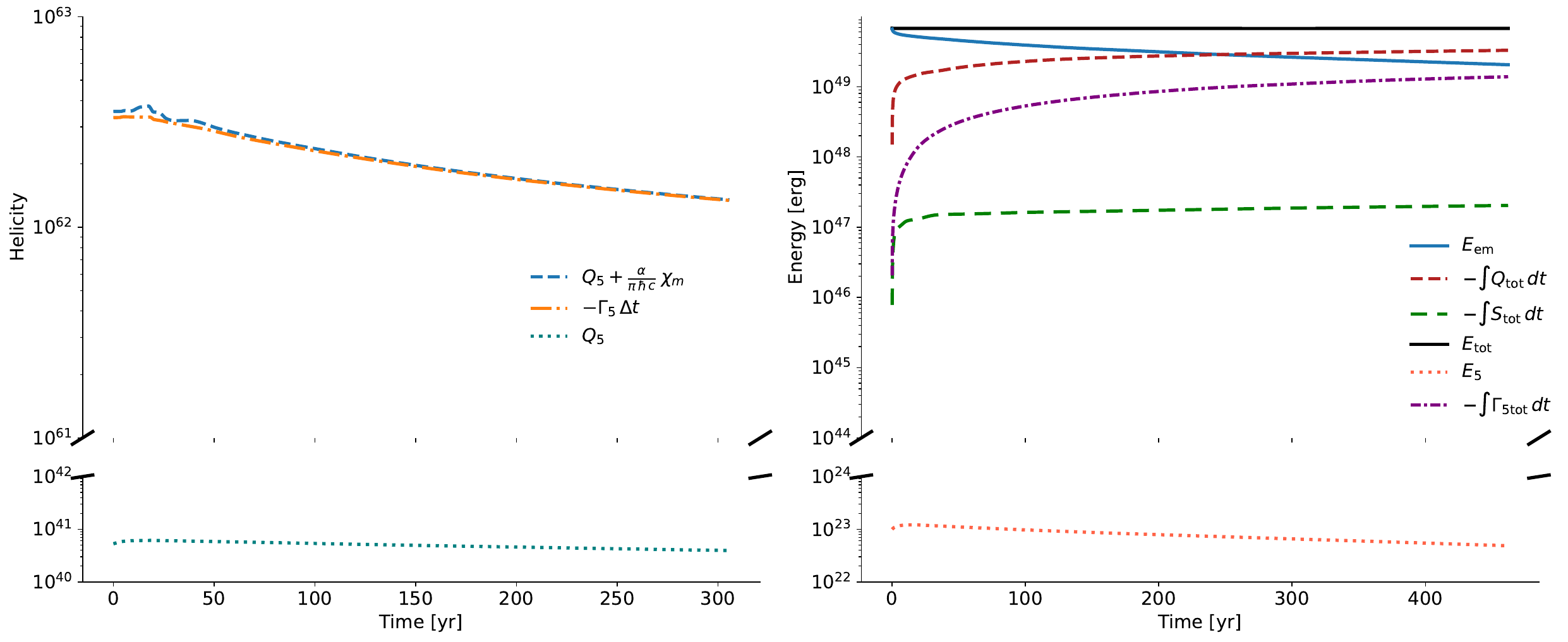}
    \caption{Generalized helicity and energy conservation. Left panel (\Eq{eq: modified helicity conservation}): Time evolution of the total helicity $Q_5 + \frac{\alpha}{\pi\,\hbar\,c}\,\chi_m$ (dashed blue), the losses due to the spin-flip term at each time step $- \Gamma_5 \, \Delta t$ (dash-dotted orange), and the resulting axial charge $Q_5$ (dotted teal). All quantities are in non-dimensional form.
   Right panel (\Eq{eq: total energy}):
Time evolution of the electromagnetic energy $E_\mathrm{em}$ (solid blue), the Joule dissipation $- \int Q_\mathrm{tot}\, dt$ (dashed red), the Poynting flux $- \int S_\mathrm{tot} \, dt$ (dashed green), the total flip term $- \int \Gamma_{5 \mathrm{tot}}\, dt$ (dash-dotted purple), the chiral energy $E_5$ (dotted orange-red), and the total energy $E_\mathrm{tot}$ (solid black).  
    }
    \label{fig: conservation}
\end{figure*}

%%%%%%%%%%%%%%%%%%%%%%%%%%%%%%%%%%
%Helicity and Energy conservation 
%%%%%%%%%%%%%%%%%%%%%%%%%%%%%%%%%%
For completeness, Figure~\ref{fig: conservation} shows the generalized helicity (\Eq{eq: modified helicity conservation}) in the left panel and the total energy conservation (\Eq{eq: total energy}) in the right panel, both for Run~F. The left panel shows the time evolution of three quantities: the total helicity $Q_5 + \frac{\alpha}{\pi\,\hbar\,c}\,\chi_m$ (dashed blue); the losses due to the spin-flip term at each time step $- \Gamma_5 \, \Delta t$ (dash-dotted orange), 
representing the number of right-handed electrons flipping to left-handed ones in the entire crust during one time step; and $Q_5$ itself (dotted teal), which reflects the total excess of right-handed electrons. The figure illustrates how the magnetic helicity generates an extremely small $Q_5$ (20 orders of magnitude smaller), which remains nearly constant. This is due to the strong damping effect of the spin-flip term. In the absence of spin-flip processes, $Q_5$ would grow until it is of the order of the magnetic helicity. Moreover, the figure confirms that our simulations respect the generalized helicity conservation law (\Eq{eq: modified helicity conservation}): the change in total helicity is closely balanced by the spin-flip term, as predicted by theory. Minor discrepancies at early times (within the first few years) arise from adjustments of the initial transient stage, and are caused by small surface helicity fluxes ($\propto \EE \times \AAA$) due to 
numerical limitations imposing boundary conditions. These surface contributions decay rapidly as the system self-adjusts over a few iterations. Although reducing the timestep improves precision, it significantly increases computational cost. Nonetheless, helicity is conserved to a satisfactory degree for the purposes of this analysis.

To further ensure the physical robustness of our results and rule out numerical artifacts, we analyze the total energy balance (\Eq{eq: total energy}). Before proceeding, we briefly examine the evolution equations for the electromagnetic energy and the energy associated with the chiral imbalance:
\begin{eqnarray}
\frac{\partial E_{em}}{\partial t}   &=&  
 - Q_{\rm tot} - \int dV \frac{\alpha \mu_5}{\pi \hbar} \EE \cdot \BB 
 - S_{\rm tot}, 
\nonumber \\
   \frac{\partial E_5}{\partial t} &=& 
     -  \Gamma_{{5}_{\rm tot}} + 
    \int dV \frac{\alpha \mu_5}{\pi \hbar} \EE \cdot \BB   ~.
    \label{eq: energy}
\end{eqnarray}
In the quasi-equilibrium regime considered here, the timescale of the spin-flip term is much shorter than that of chiral density evolution, so $\partial n_5/\partial t \rightarrow 0$ and hence $\partial E_5 /\partial t \rightarrow 0$. It follows that 
\begin{equation}
   \Gamma_{{5}_{\rm tot}} =  \int \frac{\alpha \mu_5}{\pi \hbar} \EE \cdot \BB dV  ~.
\end{equation}
The electromagnetic energy evolution then simplifies to:
\begin{equation}
 \frac{\partial E_{em}}{\partial t} = - Q_{\rm tot} - S_{\rm tot}    - \Gamma_{{5}_{\rm tot}}. 
\end{equation}
We define the total energy, $E\mathrm{tot}$, which should remain constant over time, as:
\begin{equation}
  E_\mathrm{tot}  = E_\mathrm{em} 
  - \int Q_\mathrm{tot} \, dt - \int S_\mathrm{tot} \, d t - \int {\Gamma_5}_\mathrm{tot}\, d t. 
  \label{eq: app Etot}
\end{equation}

The right panel of Figure~\ref{fig: conservation} shows the time evolution of all relevant energy components,
including electromagnetic energy ($E_\mathrm{em}$), cumulative Joule dissipation ($- \int Q_\mathrm{tot} \, dt$), net Poynting flux ($- \int S_\mathrm{tot} \, dt$), the total spin-flip dissipation ($- \int {\Gamma_5}_\mathrm{tot}\, dt$). Denoting a global (or total) wavenumber by $k_{\mathrm tot}^2 \equiv {\left(\Bnabla \times \BB\right)^2 / B^2}$,
$Q_\mathrm{tot}$ and ${\Gamma_5}_\mathrm{tot}$ can be expressed as
\begin{equation}
 Q_\mathrm{tot} =  \frac{1}{4\pi} \int dV \eta (\Bnabla \times \BB - k_5 \BB)^2  \propto 
 \left(k_{\mathrm tot}^2 + k_5^2 - 2 k_5 k_{\BB} \right) B^2,
 \label{eq: Qtot app}
\end{equation}
\begin{equation}
{\Gamma_5}_\mathrm{tot} = \frac{1}{4\pi} \int dV \eta(\Bnabla \times \BB - k_5 \BB) \cdot k_5 \BB  \propto \left( k_5 k_{\BB}   - k_5^2\right) B^2.
\end{equation}
Notably, both expressions are structurally similar and are expected to be of comparable magnitude, differing mainly due to the distinction between the global wavenumber $k_{\mathrm tot}$ and the parallel wavenumber $k_{\BB}$. The $k_5$ corrections in \EEq{eq: Qtot app} have a significant dynamical impact: they enhance dissipation when $k_5 k_B$ is negative, reduce it when positive, and can completely offset magnetic dissipation when $k_B = k_5$. Overall, the total energy $E_\mathrm{tot}$ remains conserved to within less than $1\%$ throughout the simulation, with gradual magnetic energy loss dominated by Ohmic dissipation.

\end{document}